\documentclass[aps,prl,twocolumn,showpacs,superscriptaddress,preprintnumbers]{revtex4-1}

\usepackage{graphicx} 
\usepackage{dcolumn}  
\usepackage{amsmath}
\usepackage{epsfig}
\usepackage{multirow}

\long\def\inst#1{\par\nobreak\kern 4pt\nobreak
    {\itshape #1}\par\vskip 10pt plus 3pt minus 3pt}

\graphicspath{{./figs/}}
\DeclareGraphicsExtensions{.pdf,.PDF,png,.PNG}

\usepackage{ifthen}  
\newboolean{uprightparticles}
\setboolean{uprightparticles}{false} 

\usepackage{xspace} 
\usepackage{upgreek}

\usepackage{relsize}
\def\babar{\mbox{\slshape B\kern-0.1em{\smaller A}\kern-0.1em
    B\kern-0.1em{\smaller A\kern-0.2em R}}\xspace}

\def\lhcb {\mbox{LHCb}\xspace}

\def\cleo   {\mbox{CLEO}\xspace}

\def\argus  {\mbox{ARGUS}\xspace}

\def\pep    {PEP-II}

\def\MagUp {\mbox{\em Mag\kern -0.05em Up}\xspace}

\ifthenelse{\boolean{uprightparticles}}%
{

 \def\Peta        {\ensuremath{\upeta}\xspace}

 \def\Pmu         {\ensuremath{\upmu}\xspace}

 \def\Ppi         {\ensuremath{\uppi}\xspace}                 
                  
 \def\Prho        {\ensuremath{\uprho}\xspace}

 \def\PDelta      {\ensuremath{\Delta}\xspace}                 
 \def\PXi      {\ensuremath{\Xi}\xspace}                 
 \def\PLambda      {\ensuremath{\Lambda}\xspace}                 
 \def\PSigma      {\ensuremath{\Sigma}\xspace}                 
 \def\POmega      {\ensuremath{\Omega}\xspace}                 
 \def\PUpsilon      {\ensuremath{\Upsilon}\xspace}                 
 
 \def\PB      {\ensuremath{\mathrm{B}}\xspace}                 
                  
 \def\PD      {\ensuremath{\mathrm{D}}\xspace}

 \def\PK      {\ensuremath{\mathrm{K}}\xspace}

 \def\Pc      {\ensuremath{\mathrm{c}}\xspace}                 
                  
 \def\Pe      {\ensuremath{\mathrm{e}}\xspace}

 \def\Pi      {\ensuremath{\mathrm{i}}\xspace}

 \def\Pq      {\ensuremath{\mathrm{q}}\xspace}

}
{

 \def\Peta        {\ensuremath{\eta}\xspace}

 \def\Pmu         {\ensuremath{\mu}\xspace}

 \def\Ppi         {\ensuremath{\pi}\xspace}                 
                  
 \def\Prho        {\ensuremath{\rho}\xspace}

 \mathchardef\PDelta="7101
 \mathchardef\PXi="7104
 \mathchardef\PLambda="7103
 \mathchardef\PSigma="7106
 \mathchardef\POmega="710A
 \mathchardef\PUpsilon="7107
                  
 \def\PB      {\ensuremath{B}\xspace}                 
                  
 \def\PD      {\ensuremath{D}\xspace}

 \def\PK      {\ensuremath{K}\xspace}

 \def\Pc      {\ensuremath{c}\xspace}                 
                  
 \def\Pe      {\ensuremath{e}\xspace}

 \def\Pi      {\ensuremath{i}\xspace}

 \def\Pq      {\ensuremath{q}\xspace}

}

\makeatletter
\@ifundefined{ptsize}{
\newcommand{\miniscule}{\@setfontsize\miniscule{5}{6}}
}
{
\ifcase \@ptsize \relax
  \newcommand{\miniscule}{\@setfontsize\miniscule{4}{5}}
\or
  \newcommand{\miniscule}{\@setfontsize\miniscule{5}{6}}
\or
  \newcommand{\miniscule}{\@setfontsize\miniscule{5}{6}}
\fi
}
\makeatother

\DeclareRobustCommand{\optbar}[1]{\shortstack{{\miniscule (\rule[.5ex]{1.25em}{.18mm})}
  \\ [-.7ex] $#1$}}

\def\epem       {{\ensuremath{\Pe^+\Pe^-}}\xspace}

\def\mup        {{\ensuremath{\Pmu^+}}\xspace}
\def\mun        {{\ensuremath{\Pmu^-}}\xspace} 
\def\mumu       {{\ensuremath{\Pmu^+\Pmu^-}}\xspace}

\def\quark     {{\ensuremath{\Pq}}\xspace}
\def\quarkbar  {{\ensuremath{\overline \quark}}\xspace}
\def\qqbar     {{\ensuremath{\quark\quarkbar}}\xspace}

\def\cquark    {{\ensuremath{\Pc}}\xspace}
\def\cquarkbar {{\ensuremath{\overline \cquark}}\xspace}
\def\ccbar     {{\ensuremath{\cquark\cquarkbar}}\xspace}

\def\pion   {{\ensuremath{\Ppi}}\xspace}
\def\piz    {{\ensuremath{\pion^0}}\xspace}

\def\pip    {{\ensuremath{\pion^+}}\xspace}
\def\pim    {{\ensuremath{\pion^-}}\xspace}

\def\rhomeson {{\ensuremath{\Prho}}\xspace}
\def\rhoz     {{\ensuremath{\rhomeson^0}}\xspace}

\def\kaon    {{\ensuremath{\PK}}\xspace}
  \def\Kbar    {{\kern 0.2em\overline{\kern -0.2em \PK}{}}\xspace}

\def\KorKbar    {\kern 0.18em\optbar{\kern -0.18em K}{}\xspace}

\def\Kp      {{\ensuremath{\kaon^+}}\xspace}
\def\Km      {{\ensuremath{\kaon^-}}\xspace}

\def\KpKm     {\ensuremath{\Kp \kern -0.16em \Km}\xspace}
\def\KS      {{\ensuremath{\kaon^0_{\mathrm{ \scriptscriptstyle S}}}}\xspace}

\def\Kstarb  {{\ensuremath{\Kbar{}^*}}\xspace}

\newcommand{\etapr}{\ensuremath{\Peta^{\prime}}\xspace}

  \def\Dbar    {{\kern 0.2em\overline{\kern -0.2em \PD}{}}\xspace}
\def\D       {{\ensuremath{\PD}}\xspace}

\def\DorDbar    {\kern 0.18em\optbar{\kern -0.18em D}{}\xspace}
\def\Dz      {{\ensuremath{\D^0}}\xspace}

\def\Dstarp  {{\ensuremath{\D^{*+}}}\xspace}

\def\B       {{\ensuremath{\PB}}\xspace}
\def\Bbar    {{\ensuremath{\kern 0.18em\overline{\kern -0.18em \PB}{}}}\xspace}

\def\BorBbar    {\kern 0.18em\optbar{\kern -0.18em B}{}\xspace}
\def\Bz      {{\ensuremath{\B^0}}\xspace}
\def\Bzb     {{\ensuremath{\Bbar{}^0}}\xspace}
\def\Bu      {{\ensuremath{\B^+}}\xspace}
\def\Bub     {{\ensuremath{\B^-}}\xspace}

\def\BpBm    {\ensuremath{\Bu {\kern -0.16em \Bub}}\xspace}
\def\BzBzb   {\ensuremath{\Bz {\kern -0.16em \Bzb}}\xspace}
\def\BB      {\ensuremath{\B\Bbar}\xspace}

  \def\Y#1S{\ensuremath{\PUpsilon{(#1S)}}\xspace}

\def\FourS {{\Y4S}}

\def\Lbar        {{\ensuremath{\kern 0.1em\overline{\kern -0.1em\PLambda}}}\xspace}
\def\LorLbar    {\kern 0.18em\optbar{\kern -0.18em \PLambda}{}\xspace}

\def\BF         {{\ensuremath{\mathcal{B}}}\xspace}

\def\BR         {\BF}
\def\calB         {\BF}

\def\to                 {\ensuremath{\rightarrow}\xspace}

\def\order   {{\ensuremath{\mathcal{O}}}\xspace}

\newcommand{\dm}{{\ensuremath{\Delta m}}\xspace}

\def\AT#1     {\ensuremath{A_{\mathrm{T}}^{#1}}\xspace}           

\def\C#1      {\ensuremath{\mathcal{C}_{#1}}\xspace}                       
\def\Cp#1     {\ensuremath{\mathcal{C}_{#1}^{'}}\xspace}                    
\def\Ceff#1   {\ensuremath{\mathcal{C}_{#1}^{\mathrm{(eff)}}}\xspace}        
\def\Cpeff#1  {\ensuremath{\mathcal{C}_{#1}^{'\mathrm{(eff)}}}\xspace}       
\def\Ope#1    {\ensuremath{\mathcal{O}_{#1}}\xspace}                       
\def\Opep#1   {\ensuremath{\mathcal{O}_{#1}^{'}}\xspace}                    

\newcommand{\tev}{\ifthenelse{\boolean{inbibliography}}{\ensuremath{~T\kern -0.05em eV}\xspace}{\ensuremath{\mathrm{\,Te\kern -0.1em V}}}\xspace}
\newcommand{\gev}{\ensuremath{\mathrm{\,Ge\kern -0.1em V}}\xspace}
\newcommand{\mev}{\ensuremath{\mathrm{\,Me\kern -0.1em V}}\xspace}
\newcommand{\kev}{\ensuremath{\mathrm{\,ke\kern -0.1em V}}\xspace}
\newcommand{\ev}{\ensuremath{\mathrm{\,e\kern -0.1em V}}\xspace}
\newcommand{\gevc}{\ensuremath{{\mathrm{\,Ge\kern -0.1em V\!/}c}}\xspace}
\newcommand{\mevc}{\ensuremath{{\mathrm{\,Me\kern -0.1em V\!/}c}}\xspace}
\newcommand{\gevcc}{\ensuremath{{\mathrm{\,Ge\kern -0.1em V\!/}c^2}}\xspace}
\newcommand{\gevgevcccc}{\ensuremath{{\mathrm{\,Ge\kern -0.1em V^2\!/}c^4}}\xspace}
\newcommand{\mevcc}{\ensuremath{{\mathrm{\,Me\kern -0.1em V\!/}c^2}}\xspace}

\def\invfb   {\ensuremath{\mbox{\,fb}^{-1}}\xspace}

\def\order{{\ensuremath{\mathcal{O}}}\xspace}
\newcommand{\chisq}{\ensuremath{\chi^2}\xspace}

\def\gsim{{~\raise.15em\hbox{$>$}\kern-.85em
          \lower.35em\hbox{$\sim$}~}\xspace}
\def\lsim{{~\raise.15em\hbox{$<$}\kern-.85em
          \lower.35em\hbox{$\sim$}~}\xspace}

\def\sPlot{\mbox{\em sPlot}\xspace}

\newcommand{\lum} {\ensuremath{\mathcal{L}}\xspace}

\def\EVTGEN     {\mbox{\tt EvtGen}\xspace} 

\def\geantfour  {\mbox{\tt GEANT4}\xspace} 

\def\PHOTOS     {\mbox{\tt PHOTOS}\xspace} 

\def\tell1  {TELL1\xspace}
\def\ukl1   {UKL1\xspace}

\def\usedlumi   {\ensuremath{39.3\pm0.2}}  
\def\totallumi  {\ensuremath{468.2\pm2.0}}

\def\Dmeson  {\ensuremath{D} meson}

\def\mee        {\ensuremath{m(\epem)}}
\def\mmm        {\ensuremath{m(\mumu)}}

\def\mKpi       {\ensuremath{m(\Km\pip)}}

\def\nsig       {\ensuremath{N_{\rm sig}}}
\def\nnorm      {\ensuremath{N_{\rm norm}}}

\def\DzToKPimm   {\ensuremath{\Dz\to\Km\pip\mumu}}
\def\DzToKPimumu {\DzToKPimm}

\def\DzToKPiee   {\ensuremath{\Dz\to\Km\pip\epem}}

\def\KKPiPi      {\ensuremath{\Km\Kp\pip\pim}}
\def\DzToKKPiPi  {\ensuremath{\Dz\to\KKPiPi}}

\def\KPiPiPi      {\ensuremath{\Km\pip\pip\pim}}
\def\DzToKPiPiPi  {\ensuremath{\Dz\to\KPiPiPi}}

\def\PiPiPiPi      {\ensuremath{\pim\pip\pip\pim}}
\def\DzToPiPiPiPi  {\ensuremath{\Dz\to\PiPiPiPi}}

\def\DzToKPi     {\ensuremath{\Dz\to\Km\pip}}

\def\BFDzToKPiPiPi  {\ensuremath{3.98\pm0.08\pm0.10}} 

\def\BFDzToKPi      {\ensuremath{3.89\pm0.04}} 

\def\FitBiasErrKPiee   {\ensuremath{0.2}}

\def\effKPiNoErr       {\ensuremath{27.4}}

\def\effKPiPiPiNoErr   {\ensuremath{20.1}}

\def\effKPi       {\ensuremath{\effKPiNoErr\pm0.2}}

\def\effKPiPiPi   {\ensuremath{\effKPiPiPiNoErr\pm0.2}}

\def\yieldBckgRho      {\ensuremath{0.3\pm0.2}}
\def\yieldBckgCont     {\ensuremath{9.9\pm0.9}}

\def\yieldKPieeRounded {\ensuremath{68\pm9}}
\def\yieldEtaSub       {\ensuremath{19\pm7}} 
\def\yieldPhi          {\ensuremath{3.8^{+2.7}_{-1.9}}} 

\def\yieldKPi       {\ensuremath{1\,881\,950\pm1380}} 
\def\yieldKPiPiPi   {\ensuremath{260\,870\pm520}}     

\def\sensitivity    {\ensuremath{9.7}}  

\def\dataPiz    {\ensuremath{175\pm14}} 

\def\BFKPieeRnd     {\ensuremath{4.0}}
\def\BFKPieestatRnd {\ensuremath{0.5}}
\def\BFKPieesystRnd {\ensuremath{0.2}}
\def\BFKPieebfRnd   {\ensuremath{0.1}}

\def\BFKPieeEtaRnd  {\ensuremath{1.6\pm0.6\pm0.7}}
\def\BFKPieePhi     {\ensuremath{2.2^{+1.5}_{-1.1}\pm0.6}} 

\def\BFKPieeEtaUL   {\ensuremath{3.1}}
\def\BFKPieePhiUL   {\ensuremath{0.5}}

\def\systKPieeTot     {\ensuremath{5.3}} 
\def\systKPiPiPiTot   {\ensuremath{3.6}}
\def\systBFTot        {\ensuremath{3.8}} 

\newcommand{\BABARPubYear}    {18}
\newcommand{\BABARPubNumber}  {007}
\newcommand{\SLACPubNumber} {17322}

\begin{document}
\preprint{\babar-PUB-\BABARPubYear/\BABARPubNumber} 
\preprint{SLAC-PUB-\SLACPubNumber} 

\title{{\Large \bf \boldmath Observation of the decay $D^0\rightarrow K^-\pi^+e^+e^-$}}

\author{J.~P.~Lees}
\author{V.~Poireau}
\author{V.~Tisserand}
\affiliation{Laboratoire d'Annecy-le-Vieux de Physique des Particules (LAPP), Universit\'e de Savoie, CNRS/IN2P3,  F-74941 Annecy-Le-Vieux, France}
\author{E.~Grauges}
\affiliation{Universitat de Barcelona, Facultat de Fisica, Departament ECM, E-08028 Barcelona, Spain }
\author{A.~Palano}
\affiliation{INFN Sezione di Bari and Dipartimento di Fisica, Universit\`a di Bari, I-70126 Bari, Italy }
\author{G.~Eigen}
\affiliation{University of Bergen, Institute of Physics, N-5007 Bergen, Norway }
\author{D.~N.~Brown}
\author{Yu.~G.~Kolomensky}
\affiliation{Lawrence Berkeley National Laboratory and University of California, Berkeley, California 94720, USA }
\author{M.~Fritsch}
\author{H.~Koch}
\author{T.~Schroeder}
\affiliation{Ruhr Universit\"at Bochum, Institut f\"ur Experimentalphysik 1, D-44780 Bochum, Germany }
% ...
\collaboration{The \babar\ Collaboration}
\noaffiliation

\begin{abstract}
\noindent We report the observation of the rare charm decay
$\Dz\to\Km\pip\epem$, based on $468$\invfb of $\epem$ annihilation
data collected at or close to the center-of-mass energy of the
\FourS\ resonance with the \babar\ detector at the SLAC National
Accelerator Laboratory. We find the branching fraction in the
invariant mass range $0.675< \mee < 0.875\gevcc$ of the
electron-positron pair to be $\BR(\Dz\to\Km\pip\epem) = (\BFKPieeRnd
\pm \BFKPieestatRnd \pm \BFKPieesystRnd \pm \BFKPieebfRnd) \times
10^{-6}$, where the first uncertainty is statistical, the second
systematic, and the third due to the uncertainty in the branching
fraction of the decay $\Dz\to\Km\pip\pip\pim$ used as a normalization
mode. The significance of the observation corresponds to 9.7 standard
deviations including systematic uncertainties. This result is
consistent with the recently reported $\Dz\to\Km\pip\mup\mun$
branching fraction, measured in the same invariant mass range, and
with the value expected in the Standard Model. In a set of regions of
\mee\ where long-distance effects are potentially small, we determine
a 90\% confidence level upper limit on the branching fraction
$\calB(\Dz\to\Km\pip\epem) < \BFKPieeEtaUL\times 10^{-6}$.
\end{abstract}
  

\maketitle

The decay \DzToKPiee~\cite{conjugate} is expected to be very rare in
the standard model (SM) as it cannot occur at tree
level~\cite{Glashow:1970gm}.  Short-distance contributions to the
\DzToKPiee\ branching fraction proceed through loop and box
diagrams~\cite{Paul:2011ar} and are expected to be
$\order(10^{-9})$.
However, decays with long-distance contributions,
such as $\Dz\to VX$, where $V$ is a vector or pseudoscalar meson decaying to two
leptons and $X$ is an accompanying particle or
particles, could contribute at the level of $\order(10^{-6})$ through
photon pole amplitudes or vector meson
dominance~\cite{Fajfer:2001sa,Fajfer:1998rz,Cappiello:2012vg,Paul:2011ar,Paul:2012ab}.

Certain physics models beyond the standard model, such as minimal
supersymmetric or R-parity-violating supersymmetric theories, predict
branching fractions as high as
$\order(10^{-5})$~\cite{Paul:2011ar,Paul:2012ab,Burdman:2001tf,Fajfer:2005ke,PhysRevD.76.074010}.
As virtual particles can enter in the one-loop processes, this type of
decay can be used to study new physics processes at large mass
scales. These processes could potentially be detected in regions where the decays of
intermediate mesons do not dominate.

Over the last few years there have been a number of measurements of
the decays of \B\ mesons to final states involving one or more charged
leptons. Some of these measurements suggest a possible deviation from
the assumption that all leptons couple
equally~\cite{PhysRevD.88.072012,PhysRevD.92.072014,PhysRevD.94.072007,PhysRevLett.118.211801,LHCb-PAPER-2014-024,LHCb-PAPER-2015-025,LHCb-PAPER-2017-013,LHCb-PAPER-2017-035,LHCb-PAPER-2017-017,LHCb-PAPER-2017-027}. The
possibility therefore exists that a deviation from lepton universality
will be seen in \Dmeson\ decays.

Recently, the \lhcb\ Collaboration measured $\BR(\DzToKPimm) =
(4.17\pm0.12\pm0.40)\times 10^{-6}$ in the mass range $0.675< \mmm <
0.875\gevcc$, where the decay is dominated by the \rhoz\ and $\omega$
resonances~\cite{LHCb-PAPER-2015-043}. For modes involving electrons,
the \cleo\ Collaboration set 90\% confidence level (CL) limits on the
branching fractions $\BR(\Dz\to X\ell^{+}\ell^{-})$ in the range
$(4.5-118)\times 10^{-5}$, where $X$ represents a $\piz$, $\KS$,
$\eta$, $\rhoz$, $\omega$, or $\phi$ meson and $\ell=e$ or
$\mu$~\cite{Freyberger:1996it}. The E791 Collaboration has reported
$\BR(\DzToKPiee) < 38.5\times 10^{-5}$ at the 90\% CL in the full
$m(\Km\pip)$ invariant mass range and $\BR(\DzToKPiee) < 4.7\times
10^{-5}$ in the $m(\Km\pip)$ mass range within $55\mevcc$ of the
$\Kstarb(892)^{0}$ mass~\cite{Aitala:1999db, Aitala:2000kk}.

We report here the observation of the decay \DzToKPiee~\cite{conjugate} with data
recorded with the \babar\ detector at the \pep\ asymmetric-energy
$\epem$ collider operated at the SLAC National Accelerator
Laboratory. The data sample corresponds to 424\invfb\ of
\epem\ collisions collected at the center-of-mass energy of the
\FourS\ resonance (onpeak) and an additional 44\invfb\ of data
collected 40~\mev\ below the \FourS\ resonance
(offpeak)~\cite{Lees:2013rw}. The signal branching fraction
$\BR(\DzToKPiee)$ is measured relative to the normalization decay
\DzToKPiPiPi. The \Dz\ mesons are reconstructed from the decay
$\Dstarp\to\Dz\pip$ produced in $\epem\to\ccbar$ events.  The use of
this decay chain increases the purity of the sample at the expense of
a smaller number of reconstructed \Dz\ mesons.

The \babar\ detector is described in detail in
Ref.~\cite{Aubert:2001tu,*TheBABAR:2013jta}.  Charged particles are
reconstructed as tracks with a five-layer silicon vertex detector and
a 40-layer drift chamber inside a $1.5\,$T solenoidal magnet. An
electromagnetic calorimeter comprised of 6580 CsI(Tl) crystals
is used to identify electrons and photons. A ring-imaging Cherenkov
detector is used to identify charged hadrons and to provide
additional lepton identification information.  Muons are
identified with an instrumented magnetic-flux return.

Monte Carlo (MC) simulation is used to evaluate the level of
background contamination and selection efficiencies. Simulated events
are also used to cross-check the selection procedure and for studies
of systematic effects.  The signal and normalization channels are
simulated with the \EVTGEN\ package~\cite{Lange:2001uf}. We generate
the signal channel decay with a phase-space model, while the
normalization mode includes two-body and three-body intermediate
resonances, as well as nonresonant decays. For background studies, we
generate $\epem\to\qqbar$ ($q=u,d,s,c$), dimuon, Bhabha elastic
\epem\ scattering, \BB\ background, and two-photon
events~\cite{Ward:2002qq, Sjostrand:1993yb}. The background samples
are produced with an integrated luminosity approximately six times
that of the data. Final-state radiation is provided by
\PHOTOS~\cite{Golonka:2005pn}. The detector response is simulated with
\geantfour~\cite{Agostinelli:2002hh,Allison:2006ve}. All simulated
events are reconstructed in the same manner as the data.

Events are required to contain at least five charged tracks. Candidate
\Dz\ mesons are formed from four charged tracks reconstructed with the
appropriate mass hypothesis for the \DzToKPiee\ and
\DzToKPiPiPi\ decays. Particle identification (PID) is applied to the
charged tracks and the same criteria are applied to the signal and
normalization modes~\cite{Adam:2004fq,TheBABAR:2013jta}. The four
tracks must form a good-quality vertex with a \chisq\ probability for
the vertex fit greater than 0.005. In the case of \DzToKPiee, a
bremsstrahlung energy recovery algorithm is applied to the electrons,
in which the energy of photon showers that are within a small angle
(typically 35 mrad) of the initial electron direction are added to the
energy of the electron candidate. The electron-positron pair must have
an invariant mass $\mee>0.1\gevcc$. The \Dz\ candidate momentum in the
\pep\ center-of-mass system, $p^{\ast}$, must be greater than
2.4\gevc. The requirement for five charged tracks strongly suppresses
backgrounds from QED processes. The $p^{\ast}$ criterion removes most
sources of combinatorial background and also charm hadrons produced in
\B\ decays, which are kinematically limited to less than
$\sim$2.2\gevc.

The candidate \Dstarp\ is formed by combining the \Dz\ candidate with
a charged pion with a momentum in the laboratory frame greater than
$0.1\gevc$. The pion is required to have a charge opposite to that of
the kaon in the \Dz\ decay.  A vertex fit is performed with the
\Dz\ mass constrained to its known value and the requirement that the
\Dz\ meson and the pion originate from the interaction region. The
\chisq\ probability of the fit is required to be greater than
0.005. The \Dz\ meson mass $m(\Dz)$ must be in the range
$1.81<m(\Dz)<1.91\gevcc$ and the mass difference, $\dm =
m(\Dstarp)-m(\Dz)$, between the reconstructed masses of the
\Dstarp\ and \Dz\ candidates is required to satisfy
$0.143<\dm<0.148\gevcc$. The regions around the peak positions in
$m(\Dz)$ and \dm\ in data are kept hidden until the analysis steps are
finalized.

To reject misreconstructed \DzToKPiee\ candidates that originate from
\Dz\ hadronic decays with large branching fractions where one or more
charged tracks are misidentified by the PID, the
candidate is reconstructed assuming the kaon or pion mass hypothesis
for the leptons. If the resulting candidate $m(\Dz)$ is within
20\mevcc\ of the known \Dz\ mass~\cite{PDG2018} and $|\dm|<2\mevcc$,
the event is discarded. After these criteria are applied, the
background from these hadronic decays is negligible. Multiple
candidates occur in less than $4\%$ of simulated \DzToKPiPiPi\ decays
and in less than $2\%$ of \DzToKPiee\ decays. If two or more
candidates are found in an event, the one with the highest vertex
\chisq\ probability is selected.

After the application of all selection criteria and corrections for
small differences between data and MC simulation in tracking and PID
performance, the average reconstruction efficiency for the
\DzToKPiPiPi\ decay is $\hat{\epsilon}_{\rm norm}=(\effKPiPiPi)\%$,
where the uncertainty is due to the limited size of the simulation
sample. For the \DzToKPiee\ decay, the average reconstruction
efficiency $\hat{\epsilon}_{\rm sig}$ varies between 5.0\% and 8.9\%
depending on the $\mee$ mass range. The remaining background comes
predominantly from $\epem\to\ccbar$ events. No evidence is found in MC
simulation for backgrounds that peak in the $m(\Dz)$ and \dm\ signal region.

The \DzToKPiee\ branching fraction is determined relative to that of the
normalization decay channel \DzToKPiPiPi\ using
\begin{equation}
\label{eq:ratio}
  \frac{\BR(\DzToKPiee)}{\BR(\DzToKPiPiPi)} = 
  \frac{\hat{\epsilon}_{\rm norm}}{\nnorm} \frac{\lum_{\rm norm}}{\lum_{\rm sig}}
  \sum_{i}^{\nsig} \frac{1}{\epsilon^{i}_{\rm sig}},
\end{equation}

\noindent where \hbox{$\BR(\DzToKPiPiPi)$} is the branching fraction
of the normalization mode~\cite{PDG2018}, and \nnorm\ and
$\hat{\epsilon}_{\rm norm}$ are the \hbox{\DzToKPiPiPi} fitted yield
and the reconstruction efficiency calculated from simulated
\DzToKPiPiPi\ decays, respectively. The fitted \DzToKPiee\ signal
yield is represented by \nsig, and $\epsilon^{i}_{\rm sig}$ is the
reconstruction efficiency for each signal candidate $i$, calculated
from MC simulated \DzToKPiee\ decays as a function of \mee\ and
\mKpi. The symbols $\lum_{\rm sig}$ and $\lum_{\rm norm}$ represent
the integrated luminosities used for the signal \DzToKPiee\ decay
(\totallumi\invfb) and the normalization \DzToKPiPiPi\ decay
(\usedlumi\invfb), respectively~\cite{Lees:2013rw}. The signal mode
uses both the onpeak and offpeak data samples while the normalization
mode uses only a subset of the offpeak data.

The \DzToKPiee\ and \DzToKPiPiPi\ yields are determined from extended
unbinned maximum likelihood fits to the \dm\ and the four-body mass
distributions. The \dm\ and the four-body mass distributions are not
correlated and are treated as independent observables in the fit.  For
the \DzToKPiee\ signal, a Gaussian-like function with different lower
and upper widths is used for both \dm and $m(\Km\pip\epem)$. This
asymmetric function is used in order to describe the imperfect
bremsstrahlung energy recovery for the electrons. The background in
the \DzToKPiee\ channel is modeled with an \argus\ threshold
function~\cite{Albrecht:1990am} for \dm\ and a first-order Chebyshev
polynomial for $m(\Km\pip\epem)$. For the \DzToKPiPiPi\ normalization
mode, the \dm\ and $m(\Km\pip\pim\pip)$ distributions are each
represented by two Cruijff functions with shared
means~\cite{cruijff}. The background is represented by an
\argus\ threshold function for \dm\ and a second-order Chebyshev
polynomial for $m(\Km\pip\pim\pip)$. All yields and shape parameters
are allowed to vary in the fits except for the \argus\ function
threshold end point, which is set to the kinematic threshold for the 
$\Dstarp\to\Dz\pip$ decay.

Decays of intermediate mesons to the final state $\epem\gamma$ can
potentially appear in the \mee\ spectrum as the photon is not
reconstructed.  However, the constraint $m(\Dz)>1.81\gevcc$ is
effective in reducing the background from these decays despite their
relatively high branching fractions.  We investigate the backgrounds
by generating simulation samples $\Dz\to\Km\pip V$, with intermediate
decays $\rhoz/\omega/\phi\to\epem$ and $\eta/\etapr\to\epem\gamma$. In
the simulations, QED radiative corrections are provided by
\PHOTOS~\cite{Golonka:2005pn}. The branching fractions are taken from
Ref.~\cite{PDG2018}, except for the unknown
$\calB(\Dz\to\Km\pip\eta)$, which is estimated to be $(1.8\pm0.9)\%$
from the related decay $\Dz\to\KS\piz\eta$.  After applying the
selection criteria, we expect to find \yieldBckgRho\ $\epem\gamma$
background decays in the $0.675<\mee<0.875\gevcc$ range.

The fitted yield for the \DzToKPiPiPi\ normalization data sample is
\yieldKPiPiPi. For the \DzToKPiee\ signal mode, the fitted
yield, after the subtraction of the $\epem\gamma$ background, is
$\yieldKPieeRounded$ in the range
$0.675<\mee<0.875\gevcc$. The significance $S =
\sqrt{-2\Delta\ln\lum}$ of the signal yield in this mass range,
including statistical and systematic uncertainties, is
\sensitivity\ standard deviations ($\sigma$), where $\Delta\ln\lum$ is
the change in the log-likelihood from the maximum value to the value
when the number of \DzToKPiee\ signal decays is set to $\nsig=0$.

\begin{figure}[htbp!]
\begin{center}
  \begin{tabular}{c}
  \includegraphics[width=0.97\columnwidth]{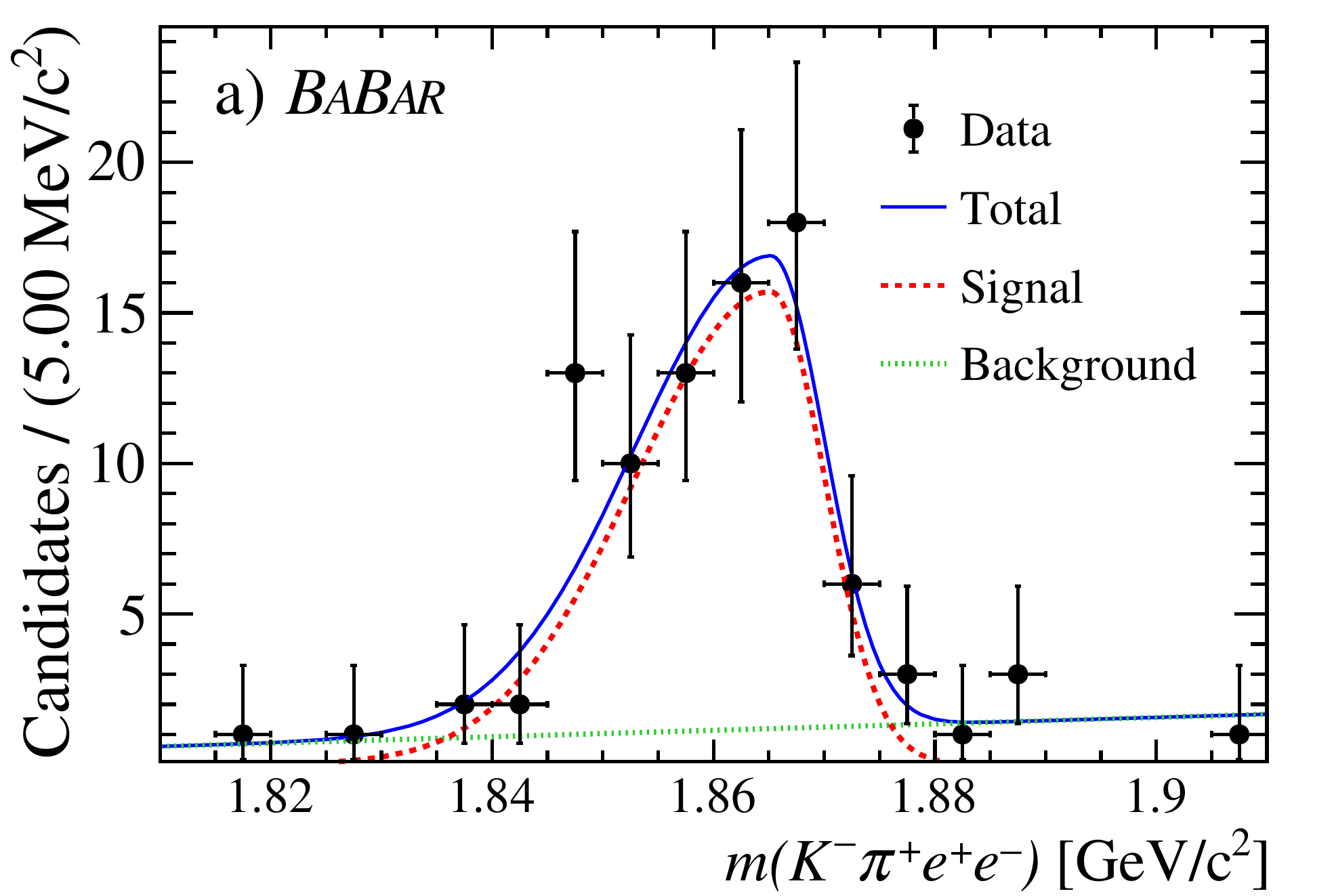} \\
  \includegraphics[width=0.97\columnwidth]{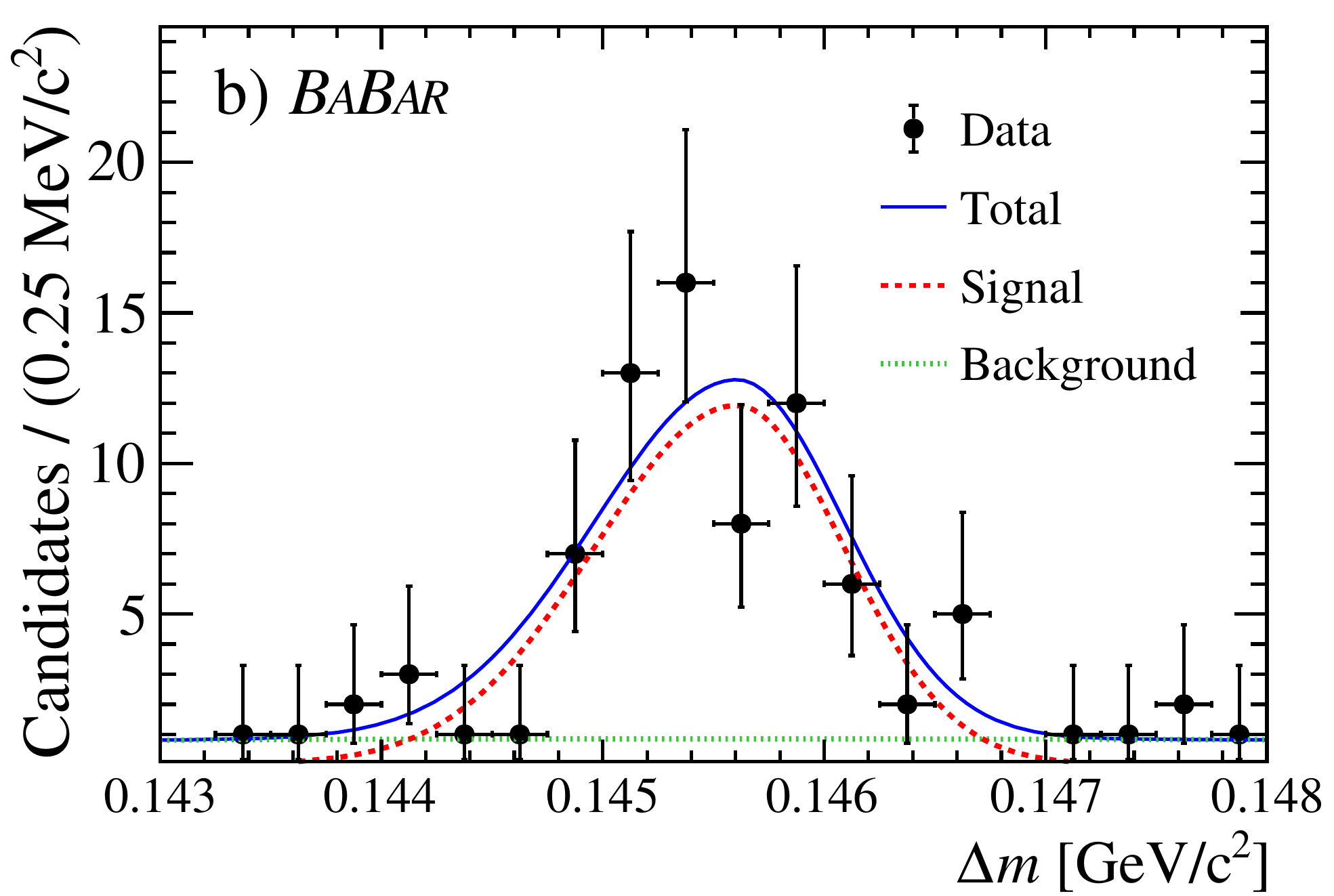} 
\end{tabular}
\end{center}
\caption{Fits to \DzToKPiee\ data distributions for (a) $m(\Km\pip\epem)$
  and (b) \dm\ mass for candidates with $0.675< \mee < 0.875\gevcc$.
}
\label{fig1}
\end{figure}

\begin{figure}[htbp!]
\begin{center}
  \begin{tabular}{c}
  \includegraphics[width=0.97\columnwidth]{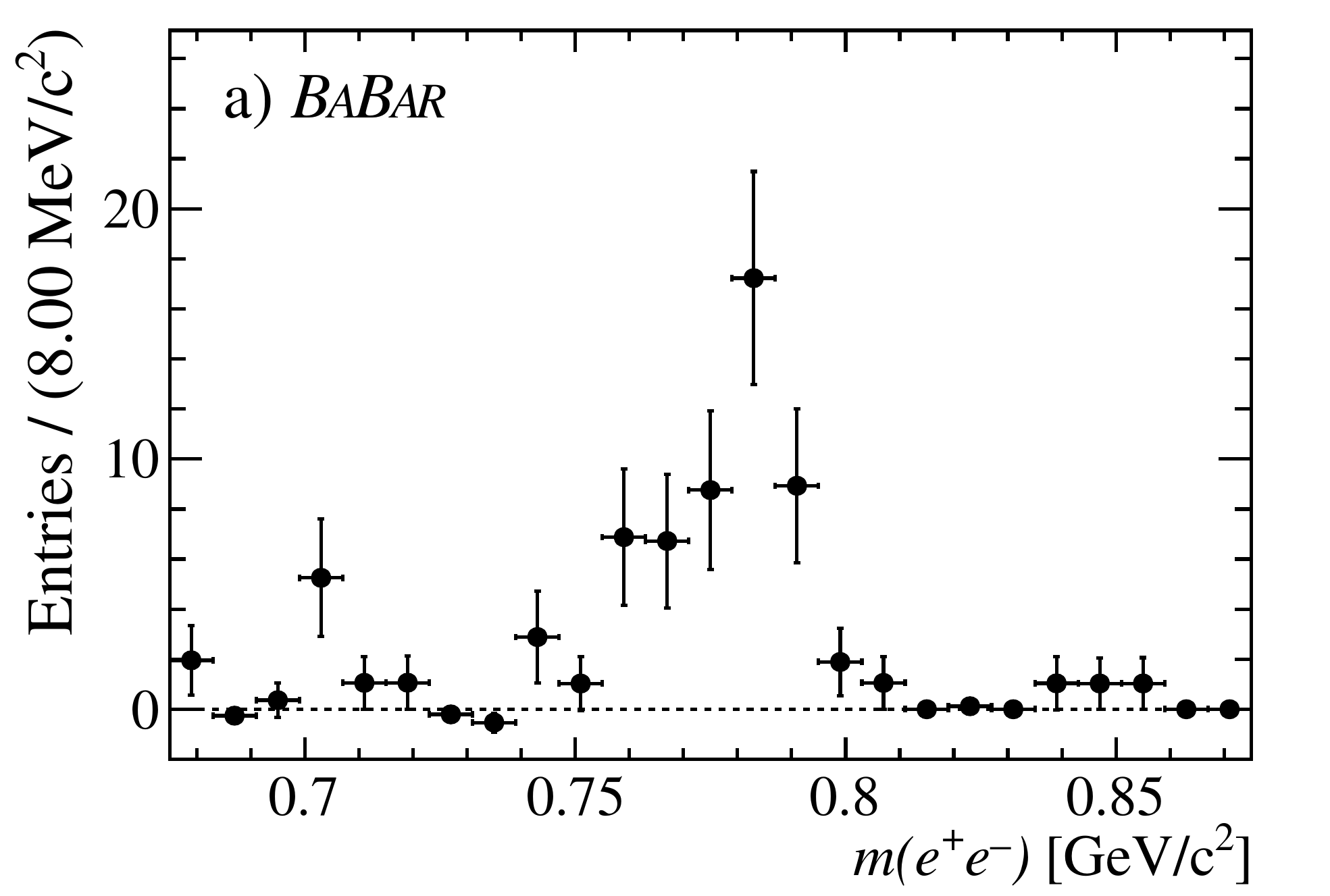} \\
  \includegraphics[width=0.97\columnwidth]{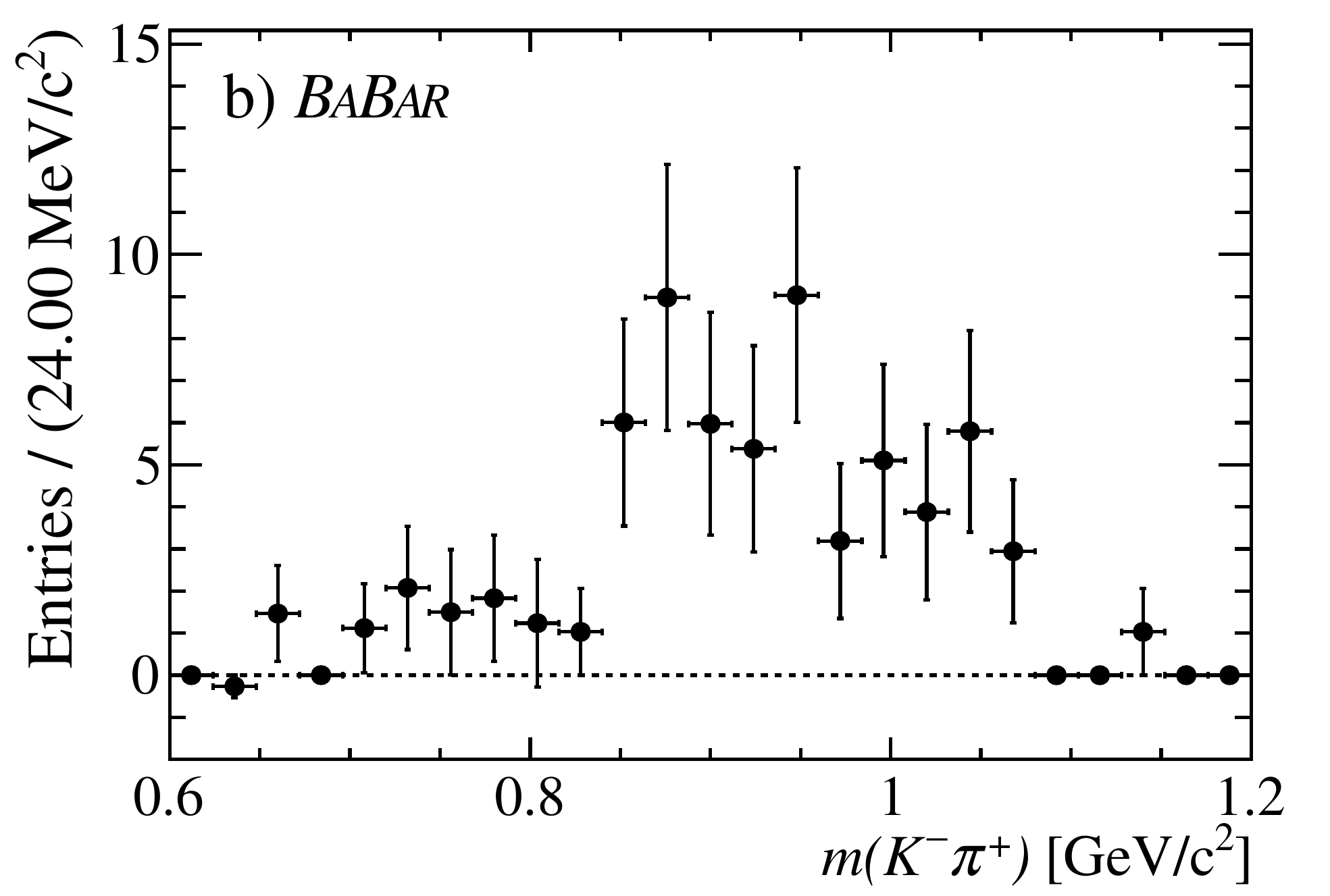} 
\end{tabular}
\end{center}
\caption{Projections of the fits to the \mbox{\DzToKPiee} data
  distributions onto (a) \mee\ and (b) \mKpi\ for candidates with
  $0.675< \mee < 0.875\gevcc$. The background has been subtracted
  using the \sPlot\ technique~\cite{Pivk:2004ty}.}
\label{fig2}
\end{figure}

Figure~\ref{fig1} shows the results of the fit to the
$m(\Km\pip\epem)$ and \dm\ distributions of the
\DzToKPiee\ signal mode in the mass range
$0.675<\mee<0.875\gevcc$. Figure~\ref{fig2} shows the projection of
the fit to the \DzToKPiee\ signal mode as a function of \mee\ and
\mKpi, where the background has been subtracted using the
\sPlot\ technique~\cite{Pivk:2004ty}. A peaking structure is visible
in \mee\ centered near the $\rhoz$ mass. A broader structure is seen
in \mKpi\ near the known mass of the $\Kstarb(892)^{0}$ meson. Both
distributions are similar to the distributions shown in
Ref.~\cite{LHCb-PAPER-2015-043} for the decay
\DzToKPimm.

\begin{figure}[htbp!]
\begin{center}
\begin{tabular}{c}
  \includegraphics[width=0.97\columnwidth]{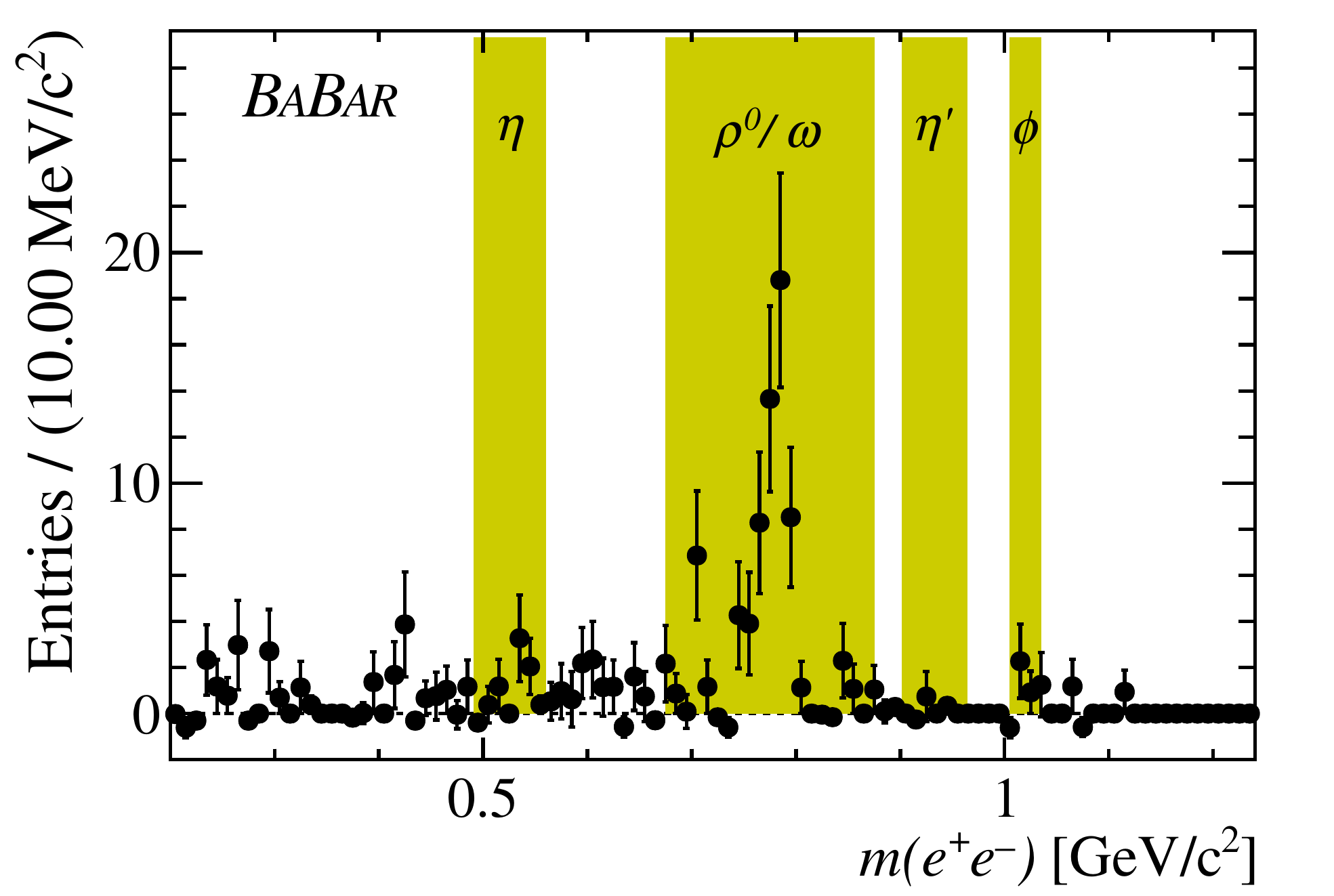}
\end{tabular}
\end{center}
\caption{Projection of the fits to the \mbox{\DzToKPiee} data
  distributions onto \mee\ for candidates with $\mee>0.2\gevcc$. The
  background has been subtracted using the
  \sPlot\ technique~\cite{Pivk:2004ty}. The shaded bands
  indicate the \mee\ regions excluded from the ``continuum'' region.}
\label{fig3}
\end{figure}

We test the performance of the maximum likelihood fit by generating
ensembles of MC simulation pseudodata samples from both the PDF
distributions and the fully simulated MC events. The mean number of
signal, normalization, and background yields used in the ensembles is
taken from the fits to the data sample. The yields are allowed to
fluctuate according to a Poisson distribution and all fit parameters
are allowed to vary. No significant bias is observed in the
normalization mode. The largest fit bias observed in the signal mode
is $0.4\pm0.1$. The biases are much smaller than the
statistical uncertainties in the yields. The fit biases are subtracted
from the fitted yields before calculating the signal branching
fractions.

To cross-check the normalization procedure, the signal mode
\DzToKPiee\ in Eq.~(\ref{eq:ratio}) is replaced with the decay
\DzToKPi, which has a well-known branching fraction~\cite{PDG2018}.
The \DzToKPi\ decay is selected using the same criteria as used for
the \DzToKPiPiPi\ mode, which is used as the normalization mode.  The
\DzToKPi\ yield is determined using an unbinned maximum likelihood fit
to \dm\ and the two-body invariant mass $m(\Km\pip)$. Three Crystal
Ball functions~\cite{Skwarnicki:1986xj} with shared means are used for
the \DzToKPi\ signal \dm\ and $m(\Km\pip)$ distributions. The
backgrounds are represented by an \argus\ function for \dm\ and a
second-order Chebyshev polynomial for $m(\Km\pip)$. The
\DzToKPi\ signal yield is \yieldKPi\ with an average
reconstruction efficiency of $\hat{\epsilon}_{\rm sig} =
(\effKPi)\%$. We determine $\BR(\DzToKPi) = (\BFDzToKPiPiPi)$\% using
Eq.~(\ref{eq:ratio}), where the uncertainties are statistical and
systematic, respectively; the current world-average is
$(\BFDzToKPi)$\%~\cite{PDG2018}. Similar compatibility with the
$\BR(\DzToKPi)$ world-average, but with larger uncertainties, is
achieved when the normalization mode \DzToKPiPiPi\ in
Eq.~(\ref{eq:ratio}) is replaced with the four-body decay modes
\DzToKKPiPi\ or \DzToPiPiPiPi.

The main sources of systematic uncertainty are associated with the
model parameterizations used in the fits and the normalization
procedure, signal MC model, fit bias, tracking and PID efficiencies,
luminosity, backgrounds from intermediate decays to $\epem\gamma$, and
the normalization mode branching fraction. Some of the tracking and
PID systematic effects cancel in the branching fraction determination
since they affect both the signal and normalization modes.

Systematic uncertainties associated with the model parameterization
are estimated by repeating the fit with the \DzToKPiee\ signal
parameters for the \dm\ and four-body distributions fixed to values
taken from simulation. Alternative fits are also performed with the
default peaking and background functions for the signal and normalization modes replaced with
alternative functions. The resulting uncertainties are 1.9\% and 1.0\%
for the signal and normalization yields, respectively.

In the mass range $0.675< \mee < 0.875\gevcc$, we replace the signal
phase-space simulation model with a model assuming
$\Dz\to\Kstarb(892)^{0}\rhoz$ with $\Kstarb(892)^{0}\to\Km\pip$ and $\rhoz\to\epem$
and assign half the difference with the default reconstruction
efficiency as a systematic uncertainty, equivalent to a relative
change of 1.8\%. We also use this number as an estimate of the
relative change in other regions of \mee\ and \mKpi\ where no suitable
alternative simulation model exists.

The systematic uncertainty in the fit bias for the signal yield is
taken from the ensemble of fits to the MC pseudodata samples and we
attribute a value of half the largest fit bias found,
$\pm\FitBiasErrKPiee$. To account for imperfect knowledge
of the tracking efficiency, we assign an uncertainty of 0.8\% per
track for the leptons and 0.7\% for the kaon and
pion~\cite{Allmendinger:2012ch}. For the PID, we estimate an
uncertainty of 0.7\% per electron, 0.2\% per pion, and 1.1\% per
kaon~\cite{TheBABAR:2013jta}. A systematic uncertainty of 0.8\% is
associated with the knowledge of the luminosity ratio, $\lum_{\rm
  norm}/\lum_{\rm sig}$~\cite{Lees:2013rw}.

The overall systematic uncertainty in the yields is \systKPieeTot\%
for the signal and \systKPiPiPiTot\% for the normalization mode. As
the PID and tracking systematic uncertainties of
the kaons and pions are correlated and cancel, the combined systematic
uncertainty in the \DzToKPiee\ branching fraction is \systBFTot\%, where the uncertainty in
the \DzToKPiPiPi\ branching fraction is excluded~\cite{PDG2018}.

The branching fraction $\BR(\DzToKPiee)$ in the mass range
$0.675<\mee<0.875\gevcc$ is determined to be $(\BFKPieeRnd \pm
\BFKPieestatRnd \pm \BFKPieesystRnd \pm \BFKPieebfRnd) \times
10^{-6}$, where the first uncertainty is statistical, the second
systematic, and the third comes from the uncertainty in
\BR(\DzToKPiPiPi)~\cite{PDG2018}. This result is compatible within the
uncertainties with $\BR(\DzToKPimumu)$ reported in
Ref.~\cite{LHCb-PAPER-2015-043}.

In the region $0.1<\mee<0.2\gevcc$, the fitted signal yield
  is \dataPiz, with the distribution dominated by the
  decay $\Dz\to\Km\pip\piz$, $\piz\to\epem\gamma$, where the photon
  has not been reconstructed.

Figure~\ref{fig3} shows the projection of the signal yield
  as a function of \mee\ for the fit to \dm\ and $m(\Km\pip\epem)$ in
  the mass range $\mee>0.2\gevcc$ above the $\piz\to\epem\gamma$ decay region, where
  the background has been subtracted using the \sPlot\ technique.
 
We determine the signal yield in the region of the $\phi$ meson by
repeating the fit to \dm\ and $m(\Km\pip\epem)$ with the
\mee\ distribution restricted to the mass range
$1.005<\mee<1.035\gevcc$. This range corresponds to $\pm 3$ times the
$\phi$ mass width, based on simulation and taking into account the
detector resolution. The fitted yield is \yieldPhi, where
the uncertainty is statistical only; the statistical significance $S$
is $1.8\sigma$. The branching fraction is determined to be
$(\BFKPieePhi)\times 10^{-7}$, where the second uncertainty is
systematic and is dominated by the uncertainty on the model
parameterization. We use the
frequentist approach of Feldman and Cousins~\cite{Feldman:1997qc} to
determine a 90\% CL branching fraction upper limit of
$\BFKPieePhiUL\times 10^{-6}$.

We repeat the fit to \dm\ and $m(\Km\pip\epem)$ in the ``continuum''
\mee\ region that is predicted to be relatively unaffected by
intermediates states, and is defined by excluding the following
\mee\ mass ranges: $\mee<0.2\gevcc$, $0.675<\mee<0.875\gevcc$,
$0.491<\mee<0.560\gevcc$, $0.902<\mee<0.964\gevcc$, and
$1.005<\mee<1.035\gevcc$. These correspond to ranges dominated by the
decays of the $\piz$ and $\rhoz/\omega$ mesons or potentially affected
by the decays of $\eta$, $\etapr$, and $\phi$ mesons, respectively.
Simulation samples of $\Dz\to\Km\pip\eta$ and $\Dz\to\Km\pip\etapr$,
with $\eta/\etapr\to\epem\gamma$, are used to determine the asymmetric
\mee\ mass ranges centered on the known $\eta$ and $\etapr$
masses. These \mee\ mass ranges exclude $90\%$ of any remaining
simulated $\eta$ and $\etapr$ candidates that pass the selection
criteria. The number of background decays from intermediate states in the
continuum region is predicted to be $\yieldBckgCont$, dominated by the
decay $\rhoz/\omega\to\epem$ with \mee\ less than $0.675\gevcc$. The fitted
yield in the continuum region, after the subtraction of this
background, is \yieldEtaSub, with a statistical
significance $S=2.6\sigma$. This corresponds to a branching fraction
$(\BFKPieeEtaRnd)\times 10^{-6}$, where the second uncertainty is
systematic and is dominated by our knowledge of the model
parameterization. The result is not significant and we determine a
90\% CL branching fraction upper limit of $\BFKPieeEtaUL\times
10^{-6}$.

In summary, we have presented the first observation of the decay
\DzToKPiee. The branching fraction in the mass range
$0.675<\mee<0.875\gevcc$ is $(\BFKPieeRnd \pm \BFKPieestatRnd \pm
\BFKPieesystRnd \pm \BFKPieebfRnd) \times 10^{-6}$, compatible with
the result for $\BR(\DzToKPimm)$~\cite{LHCb-PAPER-2015-043}, and with
theoretical predictions for the SM
contribution~\cite{Cappiello:2012vg} for this mass region. We have
placed 90\% CL branching fraction upper limits on the decay
\DzToKPiee\ in the \mee\ mass region of the $\phi$ meson and in
\mee\ mass regions where long-distance effects are potentially small.

We are grateful for the excellent luminosity and machine conditions
provided by our \pep\ colleagues, 
and for the substantial dedicated effort from
the computing organizations that support \babar.
The collaborating institutions wish to thank 
SLAC for its support and kind hospitality. 
This work is supported by
DOE
and NSF (USA),
NSERC (Canada),
CEA and
CNRS-IN2P3
(France),
BMBF and DFG
(Germany),
INFN (Italy),
FOM (The Netherlands),
NFR (Norway),
MES (Russia),
MINECO (Spain),
STFC (United Kingdom),
BSF (USA-Israel). 
Individuals have received support from the
Marie Curie EIF (European Union)
and the A.~P.~Sloan Foundation (USA).


\begin{thebibliography}{41}%
\makeatletter
\providecommand \@ifxundefined [1]{%
 \@ifx{#1\undefined}
}%
\providecommand \@ifnum [1]{%
 \ifnum #1\expandafter \@firstoftwo
 \else \expandafter \@secondoftwo
 \fi
}%
\providecommand \@ifx [1]{%
 \ifx #1\expandafter \@firstoftwo
 \else \expandafter \@secondoftwo
 \fi
}%
\providecommand \natexlab [1]{#1}%
\providecommand \enquote  [1]{``#1''}%
\providecommand \bibnamefont  [1]{#1}%
\providecommand \bibfnamefont [1]{#1}%
\providecommand \citenamefont [1]{#1}%
\providecommand \href@noop [0]{\@secondoftwo}%
\providecommand \href [0]{\begingroup \@sanitize@url \@href}%
\providecommand \@href[1]{\@@startlink{#1}\@@href}%
\providecommand \@@href[1]{\endgroup#1\@@endlink}%
\providecommand \@sanitize@url [0]{\catcode `\\12\catcode `\$12\catcode
  `\&12\catcode `\#12\catcode `\^12\catcode `\_12\catcode `\%12\relax}%
\providecommand \@@startlink[1]{}%
\providecommand \@@endlink[0]{}%
\providecommand \url  [0]{\begingroup\@sanitize@url \@url }%
\providecommand \@url [1]{\endgroup\@href {#1}{\urlprefix }}%
\providecommand \urlprefix  [0]{URL }%
\providecommand \Eprint [0]{\href }%
\providecommand \doibase [0]{http://dx.doi.org/}%
\providecommand \selectlanguage [0]{\@gobble}%
\providecommand \bibinfo  [0]{\@secondoftwo}%
\providecommand \bibfield  [0]{\@secondoftwo}%
\providecommand \translation [1]{[#1]}%
\providecommand \BibitemOpen [0]{}%
\providecommand \bibitemStop [0]{}%
\providecommand \bibitemNoStop [0]{.\EOS\space}%
\providecommand \EOS [0]{\spacefactor3000\relax}%
\providecommand \BibitemShut  [1]{\csname bibitem#1\endcsname}%
\let\auto@bib@innerbib\@empty
\bibitem [{con()}]{conjugate}%
  \BibitemOpen
  \href@noop {} {}\bibinfo {note} {Charge conjugation is implied
  throughout}\BibitemShut {NoStop}%
\bibitem [{\citenamefont {Glashow}\ \emph {et~al.}(1970)\citenamefont
  {Glashow}, \citenamefont {Iliopoulos}, and \citenamefont
  {Maiani}}]{Glashow:1970gm}%
  \BibitemOpen
  \bibfield  {author} {\bibinfo {author} {\bibfnamefont {S.~L.}\ \bibnamefont
  {Glashow}}, \bibinfo {author} {\bibfnamefont {J.}~\bibnamefont {Iliopoulos}},
   and \bibinfo {author} {\bibfnamefont {L.}~\bibnamefont {Maiani}},\ }\href
  {\doibase 10.1103/PhysRevD.2.1285} {\bibfield  {journal} {\bibinfo  {journal}
  {Phys. Rev. D}\ }\textbf {\bibinfo {volume} {2}},\ \bibinfo {pages} {1285}
  (\bibinfo {year} {1970})}\BibitemShut {NoStop}%
\bibitem [{\citenamefont {Paul}\ \emph {et~al.}(2011)\citenamefont {Paul},
  \citenamefont {Bigi}, and \citenamefont {Recksiegel}}]{Paul:2011ar}%
  \BibitemOpen
  \bibfield  {author} {\bibinfo {author} {\bibfnamefont {A.}~\bibnamefont
  {Paul}}, \bibinfo {author} {\bibfnamefont {I.~I.}\ \bibnamefont {Bigi}}, 
  and\ \bibinfo {author} {\bibfnamefont {S.}~\bibnamefont {Recksiegel}},\
  }\href {\doibase 10.1103/PhysRevD.83.114006} {\bibfield  {journal} {\bibinfo
  {journal} {Phys. Rev. D}\ }\textbf {\bibinfo {volume} {83}},\ \bibinfo
  {pages} {114006} (\bibinfo {year} {2011})}\BibitemShut {NoStop}%
\bibitem [{\citenamefont {Fajfer}\ \emph {et~al.}(2001)\citenamefont {Fajfer},
  \citenamefont {Prelov\v{s}ek}, and \citenamefont {Singer}}]{Fajfer:2001sa}%
  \BibitemOpen
  \bibfield  {author} {\bibinfo {author} {\bibfnamefont {S.}~\bibnamefont
  {Fajfer}}, \bibinfo {author} {\bibfnamefont {S.}~\bibnamefont
  {Prelov\v{s}ek}},  and \bibinfo {author} {\bibfnamefont {P.}~\bibnamefont
  {Singer}},\ }\href {\doibase 10.1103/PhysRevD.64.114009} {\bibfield
  {journal} {\bibinfo  {journal} {Phys. Rev. D}\ }\textbf {\bibinfo {volume}
  {64}},\ \bibinfo {pages} {114009} (\bibinfo {year} {2001})}\BibitemShut
  {NoStop}%
\bibitem [{\citenamefont {Fajfer}\ \emph {et~al.}(1998)\citenamefont {Fajfer},
  \citenamefont {Prelov\v{s}ek}, and \citenamefont {Singer}}]{Fajfer:1998rz}%
  \BibitemOpen
  \bibfield  {author} {\bibinfo {author} {\bibfnamefont {S.}~\bibnamefont
  {Fajfer}}, \bibinfo {author} {\bibfnamefont {S.}~\bibnamefont
  {Prelov\v{s}ek}},  and \bibinfo {author} {\bibfnamefont {P.}~\bibnamefont
  {Singer}},\ }\href {\doibase 10.1103/PhysRevD.58.094038} {\bibfield
  {journal} {\bibinfo  {journal} {Phys. Rev. D}\ }\textbf {\bibinfo {volume}
  {58}},\ \bibinfo {pages} {094038} (\bibinfo {year} {1998})}\BibitemShut
  {NoStop}%
\bibitem [{\citenamefont {Cappiello}\ \emph {et~al.}(2013)\citenamefont
  {Cappiello}, \citenamefont {Cata}, and \citenamefont
  {D'Ambrosio}}]{Cappiello:2012vg}%
  \BibitemOpen
  \bibfield  {author} {\bibinfo {author} {\bibfnamefont {L.}~\bibnamefont
  {Cappiello}}, \bibinfo {author} {\bibfnamefont {O.}~\bibnamefont {Cata}}, 
  and\ \bibinfo {author} {\bibfnamefont {G.}~\bibnamefont {D'Ambrosio}},\
  }\href {\doibase 10.1007/JHEP04(2013)135} {\bibfield  {journal} {\bibinfo
  {journal} {J. High Energy Phys.}\ }\textbf {\bibinfo {volume} {04}},\
  \bibinfo {pages} {135} (\bibinfo {year} {2013})}\BibitemShut {NoStop}%
\bibitem [{\citenamefont {Paul}\ \emph {et~al.}(2014)\citenamefont {Paul},
  \citenamefont {de~la Puente}, and \citenamefont {Bigi}}]{Paul:2012ab}%
  \BibitemOpen
  \bibfield  {author} {\bibinfo {author} {\bibfnamefont {A.}~\bibnamefont
  {Paul}}, \bibinfo {author} {\bibfnamefont {A.}~\bibnamefont {de~la Puente}},
   and \bibinfo {author} {\bibfnamefont {I.~I.}\ \bibnamefont {Bigi}},\
  }\href {\doibase 10.1103/PhysRevD.90.014035} {\bibfield  {journal} {\bibinfo
  {journal} {Phys.Rev. D}\ }\textbf {\bibinfo {volume} {90}},\ \bibinfo {pages}
  {014035} (\bibinfo {year} {2014})}\BibitemShut {NoStop}%
\bibitem [{\citenamefont {Burdman}\ \emph {et~al.}(2002)\citenamefont
  {Burdman}, \citenamefont {Golowich}, \citenamefont {Hewett}, and
  \citenamefont {Pakvasa}}]{Burdman:2001tf}%
  \BibitemOpen
  \bibfield  {author} {\bibinfo {author} {\bibfnamefont {G.}~\bibnamefont
  {Burdman}}, \bibinfo {author} {\bibfnamefont {E.}~\bibnamefont {Golowich}},
  \bibinfo {author} {\bibfnamefont {J.~A.}\ \bibnamefont {Hewett}},  and
  \bibinfo {author} {\bibfnamefont {S.}~\bibnamefont {Pakvasa}},\ }\href
  {\doibase 10.1103/PhysRevD.66.014009} {\bibfield  {journal} {\bibinfo
  {journal} {Phys. Rev. D}\ }\textbf {\bibinfo {volume} {66}},\ \bibinfo
  {pages} {014009} (\bibinfo {year} {2002})}\BibitemShut {NoStop}%
\bibitem [{\citenamefont {Fajfer} and \citenamefont
  {Prelov\v{s}ek}(2006)}]{Fajfer:2005ke}%
  \BibitemOpen
  \bibfield  {author} {\bibinfo {author} {\bibfnamefont {S.}~\bibnamefont
  {Fajfer}} and \bibinfo {author} {\bibfnamefont {S.}~\bibnamefont
  {Prelov\v{s}ek}},\ }\href {\doibase 10.1103/PhysRevD.73.054026} {\bibfield
  {journal} {\bibinfo  {journal} {Phys. Rev. D}\ }\textbf {\bibinfo {volume}
  {73}},\ \bibinfo {pages} {054026} (\bibinfo {year} {2006})}\BibitemShut
  {NoStop}%
\bibitem [{\citenamefont {Fajfer}\ \emph {et~al.}(2007)\citenamefont {Fajfer},
  \citenamefont {Ko\v{s}nik}, and \citenamefont
  {Prelov\v{s}ek}}]{PhysRevD.76.074010}%
  \BibitemOpen
  \bibfield  {author} {\bibinfo {author} {\bibfnamefont {S.}~\bibnamefont
  {Fajfer}}, \bibinfo {author} {\bibfnamefont {N.}~\bibnamefont {Ko\v{s}nik}},
   and \bibinfo {author} {\bibfnamefont {S.}~\bibnamefont {Prelov\v{s}ek}},\
  }\href {\doibase 10.1103/PhysRevD.76.074010} {\bibfield  {journal} {\bibinfo
  {journal} {Phys. Rev. D}\ }\textbf {\bibinfo {volume} {76}},\ \bibinfo
  {pages} {074010} (\bibinfo {year} {2007})}\BibitemShut {NoStop}%
\bibitem [{\citenamefont {Lees}\ \emph
  {et~al.}(2013{\natexlab{a}})\citenamefont {Lees} \emph
  {et~al.}}]{PhysRevD.88.072012}%
  \BibitemOpen
  \bibfield  {author} {\bibinfo {author} {\bibfnamefont {J.~P.}\ \bibnamefont
  {Lees}} \emph {et~al.} (\bibinfo {collaboration} {\babar\ Collaboration}),\
  }\href {\doibase 10.1103/PhysRevD.88.072012} {\bibfield  {journal} {\bibinfo
  {journal} {Phys. Rev. D}\ }\textbf {\bibinfo {volume} {88}},\ \bibinfo
  {pages} {072012} (\bibinfo {year} {2013}{\natexlab{a}})}\BibitemShut
  {NoStop}%
\bibitem [{\citenamefont {Huschle}\ \emph {et~al.}(2015)\citenamefont {Huschle}
  \emph {et~al.}}]{PhysRevD.92.072014}%
  \BibitemOpen
  \bibfield  {author} {\bibinfo {author} {\bibfnamefont {M.}~\bibnamefont
  {Huschle}} \emph {et~al.} (\bibinfo {collaboration} {Belle Collaboration}),\
  }\href {\doibase 10.1103/PhysRevD.92.072014} {\bibfield  {journal} {\bibinfo
  {journal} {Phys. Rev. D}\ }\textbf {\bibinfo {volume} {92}},\ \bibinfo
  {pages} {072014} (\bibinfo {year} {2015})}\BibitemShut {NoStop}%
\bibitem [{\citenamefont {Sato}\ \emph {et~al.}(2016)\citenamefont {Sato} \emph
  {et~al.}}]{PhysRevD.94.072007}%
  \BibitemOpen
  \bibfield  {author} {\bibinfo {author} {\bibfnamefont {Y.}~\bibnamefont
  {Sato}} \emph {et~al.} (\bibinfo {collaboration} {Belle Collaboration}),\
  }\href {\doibase 10.1103/PhysRevD.94.072007} {\bibfield  {journal} {\bibinfo
  {journal} {Phys. Rev. D}\ }\textbf {\bibinfo {volume} {94}},\ \bibinfo
  {pages} {072007} (\bibinfo {year} {2016})}\BibitemShut {NoStop}%
\bibitem [{\citenamefont {Hirose}\ \emph {et~al.}(2017)\citenamefont {Hirose}
  \emph {et~al.}}]{PhysRevLett.118.211801}%
  \BibitemOpen
  \bibfield  {author} {\bibinfo {author} {\bibfnamefont {S.}~\bibnamefont
  {Hirose}} \emph {et~al.} (\bibinfo {collaboration} {Belle Collaboration}),\
  }\href {\doibase 10.1103/PhysRevLett.118.211801} {\bibfield  {journal}
  {\bibinfo  {journal} {Phys. Rev. Lett.}\ }\textbf {\bibinfo {volume} {118}},\
  \bibinfo {pages} {211801} (\bibinfo {year} {2017})}\BibitemShut {NoStop}%
\bibitem [{\citenamefont {Aaij}\ \emph {et~al.}(2014)\citenamefont {Aaij} \emph
  {et~al.}}]{LHCb-PAPER-2014-024}%
  \BibitemOpen
  \bibfield  {author} {\bibinfo {author} {\bibfnamefont {R.}~\bibnamefont
  {Aaij}} \emph {et~al.} (\bibinfo {collaboration} {LHCb Collaboration}),\
  }\href {\doibase 10.1103/PhysRevLett.113.151601} {\bibfield  {journal}
  {\bibinfo  {journal} {Phys. Rev. Lett.}\ }\textbf {\bibinfo {volume} {113}},\
  \bibinfo {pages} {151601} (\bibinfo {year} {2014})}\BibitemShut {NoStop}%
\bibitem [{\citenamefont {Aaij}\ \emph {et~al.}(2015)\citenamefont {Aaij} \emph
  {et~al.}}]{LHCb-PAPER-2015-025}%
  \BibitemOpen
  \bibfield  {author} {\bibinfo {author} {\bibfnamefont {R.}~\bibnamefont
  {Aaij}} \emph {et~al.} (\bibinfo {collaboration} {LHCb Collaboration}),\
  }\href {\doibase 10.1103/PhysRevLett.115.111803} {\bibfield  {journal}
  {\bibinfo  {journal} {Phys. Rev. Lett.}\ }\textbf {\bibinfo {volume} {115}},\
  \bibinfo {pages} {111803} (\bibinfo {year} {2015})}\BibitemShut {NoStop}%
\bibitem [{\citenamefont {Aaij}\ \emph {et~al.}(2017)\citenamefont {Aaij} \emph
  {et~al.}}]{LHCb-PAPER-2017-013}%
  \BibitemOpen
  \bibfield  {author} {\bibinfo {author} {\bibfnamefont {R.}~\bibnamefont
  {Aaij}} \emph {et~al.} (\bibinfo {collaboration} {LHCb Collaboration}),\
  }\href {\doibase 10.1007/JHEP08(2017)055} {\bibfield  {journal} {\bibinfo
  {journal} {J. High Energy Phys.}\ }\textbf {\bibinfo {volume} {08}},\
  \bibinfo {pages} {055} (\bibinfo {year} {2017})}\BibitemShut {NoStop}%
\bibitem [{\citenamefont {Aaij}\ \emph
  {et~al.}(2018{\natexlab{a}})\citenamefont {Aaij} \emph
  {et~al.}}]{LHCb-PAPER-2017-035}%
  \BibitemOpen
  \bibfield  {author} {\bibinfo {author} {\bibfnamefont {R.}~\bibnamefont
  {Aaij}} \emph {et~al.} (\bibinfo {collaboration} {LHCb Collaboration}),\
  }\href {\doibase 10.1103/PhysRevLett.120.121801} {\bibfield  {journal}
  {\bibinfo  {journal} {Phys. Rev. Lett.}\ }\textbf {\bibinfo {volume} {120}},\
  \bibinfo {pages} {121801} (\bibinfo {year} {2018}{\natexlab{a}})}\BibitemShut
  {NoStop}%
\bibitem [{\citenamefont {Aaij}\ \emph
  {et~al.}(2018{\natexlab{b}})\citenamefont {Aaij} \emph
  {et~al.}}]{LHCb-PAPER-2017-017}%
  \BibitemOpen
  \bibfield  {author} {\bibinfo {author} {\bibfnamefont {R.}~\bibnamefont
  {Aaij}} \emph {et~al.} (\bibinfo {collaboration} {LHCb Collaboration}),\
  }\href {\doibase 10.1103/PhysRevLett.120.171802} {\bibfield  {journal}
  {\bibinfo  {journal} {Phys. Rev. Lett.}\ }\textbf {\bibinfo {volume} {120}},\
  \bibinfo {pages} {171802} (\bibinfo {year} {2018}{\natexlab{b}})}\BibitemShut
  {NoStop}%
\bibitem [{\citenamefont {Aaij}\ \emph
  {et~al.}(2018{\natexlab{c}})\citenamefont {Aaij} \emph
  {et~al.}}]{LHCb-PAPER-2017-027}%
  \BibitemOpen
  \bibfield  {author} {\bibinfo {author} {\bibfnamefont {R.}~\bibnamefont
  {Aaij}} \emph {et~al.} (\bibinfo {collaboration} {LHCb Collaboration}),\
  }\href {\doibase 10.1103/PhysRevD.97.072013} {\bibfield  {journal} {\bibinfo
  {journal} {Phys. Rev. D}\ }\textbf {\bibinfo {volume} {97}},\ \bibinfo
  {pages} {072013} (\bibinfo {year} {2018}{\natexlab{c}})}\BibitemShut
  {NoStop}%
\bibitem [{\citenamefont {Aaij}\ \emph {et~al.}(2016)\citenamefont {Aaij} \emph
  {et~al.}}]{LHCb-PAPER-2015-043}%
  \BibitemOpen
  \bibfield  {author} {\bibinfo {author} {\bibfnamefont {R.}~\bibnamefont
  {Aaij}} \emph {et~al.} (\bibinfo {collaboration} {LHCb Collaboration}),\
  }\href {\doibase 10.1016/j.physletb.2016.04.029} {\bibfield  {journal}
  {\bibinfo  {journal} {Phys. Lett. B}\ }\textbf {\bibinfo {volume} {757}},\
  \bibinfo {pages} {558} (\bibinfo {year} {2016})}\BibitemShut {NoStop}%
\bibitem [{\citenamefont {Freyberger}\ \emph {et~al.}(1996)\citenamefont
  {Freyberger} \emph {et~al.}}]{Freyberger:1996it}%
  \BibitemOpen
  \bibfield  {author} {\bibinfo {author} {\bibfnamefont {A.}~\bibnamefont
  {Freyberger}} \emph {et~al.} (\bibinfo {collaboration} {CLEO
  Collaboration}),\ }\href {\doibase 10.1103/PhysRevLett.76.3065} {\bibfield
  {journal} {\bibinfo  {journal} {Phys. Rev. Lett.}\ }\textbf {\bibinfo
  {volume} {76}},\ \bibinfo {pages} {3065} (\bibinfo {year}
  {1996})}\BibitemShut {NoStop}%
\bibitem [{\citenamefont {Aitala}\ \emph {et~al.}(1999)\citenamefont {Aitala}
  \emph {et~al.}}]{Aitala:1999db}%
  \BibitemOpen
  \bibfield  {author} {\bibinfo {author} {\bibfnamefont {E.~M.}\ \bibnamefont
  {Aitala}} \emph {et~al.} (\bibinfo {collaboration} {E791 Collaboration}),\
  }\href {\doibase 10.1016/S0370-2693(99)00902-8} {\bibfield  {journal}
  {\bibinfo  {journal} {Phys. Lett. B}\ }\textbf {\bibinfo {volume} {462}},\
  \bibinfo {pages} {401} (\bibinfo {year} {1999})}\BibitemShut {NoStop}%
\bibitem [{\citenamefont {Aitala}\ \emph {et~al.}(2001)\citenamefont {Aitala}
  \emph {et~al.}}]{Aitala:2000kk}%
  \BibitemOpen
  \bibfield  {author} {\bibinfo {author} {\bibfnamefont {E.~M.}\ \bibnamefont
  {Aitala}} \emph {et~al.} (\bibinfo {collaboration} {E791 Collaboration}),\
  }\href {\doibase 10.1103/PhysRevLett.86.3969} {\bibfield  {journal} {\bibinfo
   {journal} {Phys. Rev. Lett.}\ }\textbf {\bibinfo {volume} {86}},\ \bibinfo
  {pages} {3969} (\bibinfo {year} {2001})}\BibitemShut {NoStop}%
\bibitem [{\citenamefont {Lees}\ \emph
  {et~al.}(2013{\natexlab{b}})\citenamefont {Lees} \emph
  {et~al.}}]{Lees:2013rw}%
  \BibitemOpen
  \bibfield  {author} {\bibinfo {author} {\bibfnamefont {J.~P.}\ \bibnamefont
  {Lees}} \emph {et~al.} (\bibinfo {collaboration} {\babar\ Collaboration}),\
  }\href {\doibase 10.1016/j.nima.2013.04.029} {\bibfield  {journal} {\bibinfo
  {journal} {Nucl. Instrum. Methods Phys. Res., Sect A}\ }\textbf {\bibinfo
  {volume} {726}},\ \bibinfo {pages} {203} (\bibinfo {year}
  {2013}{\natexlab{b}})}\BibitemShut {NoStop}%
\bibitem [{\citenamefont {Aubert}\ \emph {et~al.}(2002)\citenamefont {Aubert}
  \emph {et~al.}}]{Aubert:2001tu}%
  \BibitemOpen
  \bibfield  {author} {\bibinfo {author} {\bibfnamefont {B.}~\bibnamefont
  {Aubert}} \emph {et~al.} (\bibinfo {collaboration} {\babar\ Collaboration}),\
  }\href {\doibase 10.1016/S0168-9002(01)02012-5} {\bibfield  {journal}
  {\bibinfo  {journal} {Nucl. Instrum. Methods Phys. Res., Sect A}\ }\textbf
  {\bibinfo {volume} {479}},\ \bibinfo {pages} {1} (\bibinfo {year}
  {2002})}\BibitemShut {NoStop}%
\bibitem [{\citenamefont {Aubert}\ \emph {et~al.}(2013)\citenamefont {Aubert}
  \emph {et~al.}}]{TheBABAR:2013jta}%
  \BibitemOpen
  \bibfield  {author} {\bibinfo {author} {\bibfnamefont {B.}~\bibnamefont
  {Aubert}} \emph {et~al.} (\bibinfo {collaboration} {\babar\ Collaboration}),\
  }\href {\doibase 10.1016/j.nima.2013.05.107} {\bibfield  {journal} {\bibinfo
  {journal} {Nucl. Instrum. Methods Phys. Res., Sect A}\ }\textbf {\bibinfo
  {volume} {729}},\ \bibinfo {pages} {615} (\bibinfo {year}
  {2013})}\BibitemShut {NoStop}%
\bibitem [{\citenamefont {Lange}(2001)}]{Lange:2001uf}%
  \BibitemOpen
  \bibfield  {author} {\bibinfo {author} {\bibfnamefont {D.~J.}\ \bibnamefont
  {Lange}},\ }\href {\doibase 10.1016/S0168-9002(01)00089-4} {\bibfield
  {journal} {\bibinfo  {journal} {Nucl. Instrum. Methods Phys. Res., Sect A}\
  }\textbf {\bibinfo {volume} {462}},\ \bibinfo {pages} {152} (\bibinfo {year}
  {2001})}\BibitemShut {NoStop}%
\bibitem [{\citenamefont {Ward}\ \emph {et~al.}(2003)\citenamefont {Ward},
  \citenamefont {Jadach}, and \citenamefont {Was}}]{Ward:2002qq}%
  \BibitemOpen
  \bibfield  {author} {\bibinfo {author} {\bibfnamefont {B.~F.~L.}\
  \bibnamefont {Ward}}, \bibinfo {author} {\bibfnamefont {S.}~\bibnamefont
  {Jadach}},  and \bibinfo {author} {\bibfnamefont {Z.}~\bibnamefont {Was}},\
  }\href {\doibase 10.1016/S0920-5632(03)80147-0} {\bibfield  {journal}
  {\bibinfo  {journal} {Nucl. Phys. Proc. Suppl.}\ }\textbf {\bibinfo {volume}
  {116}},\ \bibinfo {pages} {73} (\bibinfo {year} {2003})}\BibitemShut
  {NoStop}%
\bibitem [{\citenamefont {{T. Sj\"{o}strand}}(1994)}]{Sjostrand:1993yb}%
  \BibitemOpen
  \bibfield  {author} {\bibinfo {author} {\bibnamefont {{T. Sj\"{o}strand}}},\
  }\href {\doibase 10.1016/0010-4655(94)90132-5} {\bibfield  {journal}
  {\bibinfo  {journal} {Comput. Phys. Commun.}\ }\textbf {\bibinfo {volume}
  {82}},\ \bibinfo {pages} {74} (\bibinfo {year} {1994})}\BibitemShut {NoStop}%
\bibitem [{\citenamefont {Golonka} and \citenamefont
  {Was}(2006)}]{Golonka:2005pn}%
  \BibitemOpen
  \bibfield  {author} {\bibinfo {author} {\bibfnamefont {P.}~\bibnamefont
  {Golonka}} and \bibinfo {author} {\bibfnamefont {Z.}~\bibnamefont {Was}},\
  }\href {\doibase 10.1140/epjc/s2005-02396-4} {\bibfield  {journal} {\bibinfo
  {journal} {Eur. Phys. J. C}\ }\textbf {\bibinfo {volume} {45}},\ \bibinfo
  {pages} {97} (\bibinfo {year} {2006})}\BibitemShut {NoStop}%
\bibitem [{\citenamefont {Agostinelli}\ \emph {et~al.}(2003)\citenamefont
  {Agostinelli} \emph {et~al.}}]{Agostinelli:2002hh}%
  \BibitemOpen
  \bibfield  {author} {\bibinfo {author} {\bibfnamefont {S.}~\bibnamefont
  {Agostinelli}} \emph {et~al.} (\bibinfo {collaboration} {\geantfour\
  Collaboration}),\ }\href {\doibase 10.1016/S0168-9002(03)01368-8} {\bibfield
  {journal} {\bibinfo  {journal} {Nucl. Instrum. Methods Phys. Res., Sect A}\
  }\textbf {\bibinfo {volume} {506}},\ \bibinfo {pages} {250} (\bibinfo {year}
  {2003})}\BibitemShut {NoStop}%
\bibitem [{\citenamefont {Allison}\ \emph {et~al.}(2006)\citenamefont
  {Allison}, \citenamefont {Amako}, \citenamefont {Apostolakis}, \citenamefont
  {Araujo}, \citenamefont {Dubois} \emph {et~al.}}]{Allison:2006ve}%
  \BibitemOpen
  \bibfield  {author} {\bibinfo {author} {\bibfnamefont {J.}~\bibnamefont
  {Allison}}, \bibinfo {author} {\bibfnamefont {K.}~\bibnamefont {Amako}},
  \bibinfo {author} {\bibfnamefont {J.}~\bibnamefont {Apostolakis}}, \bibinfo
  {author} {\bibfnamefont {H.}~\bibnamefont {Araujo}}, \bibinfo {author}
  {\bibfnamefont {P.}~\bibnamefont {Dubois}},  \emph {et~al.} (\bibinfo
  {collaboration} {\geantfour\ Collaboration}),\ }\href {\doibase
  10.1109/TNS.2006.869826} {\bibfield  {journal} {\bibinfo  {journal} {IEEE
  Trans. Nucl. Sci.}\ }\textbf {\bibinfo {volume} {53}},\ \bibinfo {pages}
  {270} (\bibinfo {year} {2006})}\BibitemShut {NoStop}%
\bibitem [{\citenamefont {Adam}\ \emph {et~al.}(2005)\citenamefont {Adam} \emph
  {et~al.}}]{Adam:2004fq}%
  \BibitemOpen
  \bibfield  {author} {\bibinfo {author} {\bibfnamefont {I.}~\bibnamefont
  {Adam}} \emph {et~al.},\ }\href {\doibase 10.1016/j.nima.2004.08.129}
  {\bibfield  {journal} {\bibinfo  {journal} {Nucl. Instrum. Methods Phys.
  Res., Sect A}\ }\textbf {\bibinfo {volume} {538}},\ \bibinfo {pages} {281}
  (\bibinfo {year} {2005})}\BibitemShut {NoStop}%
\bibitem [{\citenamefont {Tanabashi}\ \emph {et~al.}(2018)\citenamefont
  {Tanabashi} \emph {et~al.}}]{PDG2018}%
  \BibitemOpen
  \bibfield  {author} {\bibinfo {author} {\bibfnamefont {M.}~\bibnamefont
  {Tanabashi}} \emph {et~al.} (\bibinfo {collaboration} {Particle Data
  Group}),\ }\href@noop {} {\bibfield  {journal} {\bibinfo  {journal} {Phys.
  Rev. D}\ }\textbf {\bibinfo {volume} {98}},\ \bibinfo {pages} {030001}
  (\bibinfo {year} {2018})}\BibitemShut {NoStop}%
\bibitem [{\citenamefont {Albrecht}\ \emph {et~al.}(1990)\citenamefont
  {Albrecht} \emph {et~al.}}]{Albrecht:1990am}%
  \BibitemOpen
  \bibfield  {author} {\bibinfo {author} {\bibfnamefont {H.}~\bibnamefont
  {Albrecht}} \emph {et~al.} (\bibinfo {collaboration} {ARGUS Collaboration}),\
  }\href {\doibase 10.1016/0370-2693(90)91293-K} {\bibfield  {journal}
  {\bibinfo  {journal} {Phys. Lett. B}\ }\textbf {\bibinfo {volume} {241}},\
  \bibinfo {pages} {278} (\bibinfo {year} {1990})}\BibitemShut {NoStop}%
\bibitem [{cru()}]{cruijff}%
  \BibitemOpen
  \href@noop {} {}\bibinfo {note} {The Cruijff function is a centered Gaussian
  with different left-right resolutions and non-Gaussian tails: $f(x) =
  \exp(-(x-m)^2/(2\sigma^2_{\rm L,R} + \alpha_{\rm L,R}(x-m)^2))$}\BibitemShut
  {NoStop}%
\bibitem [{\citenamefont {Pivk} and \citenamefont
  {Le~Diberder}(2005)}]{Pivk:2004ty}%
  \BibitemOpen
  \bibfield  {author} {\bibinfo {author} {\bibfnamefont {M.}~\bibnamefont
  {Pivk}} and \bibinfo {author} {\bibfnamefont {F.~R.}\ \bibnamefont
  {Le~Diberder}},\ }\href {\doibase 10.1016/j.nima.2005.08.106} {\bibfield
  {journal} {\bibinfo  {journal} {Nucl. Instrum. Methods Phys. Res., Sect A}\
  }\textbf {\bibinfo {volume} {555}},\ \bibinfo {pages} {356} (\bibinfo {year}
  {2005})}\BibitemShut {NoStop}%
\bibitem [{\citenamefont {Skwarnicki}(1986)}]{Skwarnicki:1986xj}%
  \BibitemOpen
  \bibfield  {author} {\bibinfo {author} {\bibfnamefont {T.}~\bibnamefont
  {Skwarnicki}},\ }\emph {\bibinfo {title} {{A study of the radiative cascade
  transitions between the $\Upsilon'$ and $\Upsilon$ resonances}}},\ \href@noop
  {} {Ph.D. thesis},\ \bibinfo  {school} {Institute of Nuclear Physics, Krakow}
  (\bibinfo {year} {1986}),\ \bibinfo {note}
  {{\href{http://inspirehep.net/record/230779/}{DESY-F31-86-02}}}\BibitemShut
  {NoStop}%
\bibitem [{\citenamefont {Allmendinger}\ \emph {et~al.}(2013)\citenamefont
  {Allmendinger} \emph {et~al.}}]{Allmendinger:2012ch}%
  \BibitemOpen
  \bibfield  {author} {\bibinfo {author} {\bibfnamefont {T.}~\bibnamefont
  {Allmendinger}} \emph {et~al.},\ }\href {\doibase 10.1016/j.nima.2012.11.184}
  {\bibfield  {journal} {\bibinfo  {journal} {Nucl. Instrum. Methods Phys.
  Res., Sect A}\ }\textbf {\bibinfo {volume} {704}},\ \bibinfo {pages} {44}
  (\bibinfo {year} {2013})}\BibitemShut {NoStop}%
\bibitem [{\citenamefont {Feldman} and \citenamefont
  {Cousins}(1998)}]{Feldman:1997qc}%
  \BibitemOpen
  \bibfield  {author} {\bibinfo {author} {\bibfnamefont {G.~J.}\ \bibnamefont
  {Feldman}} and \bibinfo {author} {\bibfnamefont {R.~D.}\ \bibnamefont
  {Cousins}},\ }\href {\doibase 10.1103/PhysRevD.57.3873} {\bibfield  {journal}
  {\bibinfo  {journal} {Phys. Rev. D}\ }\textbf {\bibinfo {volume} {57}},\
  \bibinfo {pages} {3873} (\bibinfo {year} {1998})}\BibitemShut {NoStop}%
\end{thebibliography}
\end{document}